\documentclass[twocolumn,prb,showpacs,preprintnumbers,amsmath,amssymb,floatfix]{revtex4-1}

\usepackage{graphicx}
\usepackage{dcolumn}
\usepackage{bm}
\usepackage{subfigure}

\newcommand{\EG}{E_{\Gamma}}
\newcommand{\Ek}{E_{\mathrm{kin}}}
\newcommand{\lambdaHB}{\lambda_{\rm HB}}
\newcommand{\VHB}{W_{\rm HB}}
\newcommand{\lHB}{l_{\rm HB}}
\newcommand{\kHB}{k_{\rm HB}}

\begin{document}

\title{
   Diffractive wave guiding of hot electrons by the Au (111) herringbone reconstruction }
\author{ F. Libisch$^{1,2}$, V. Geringer$^3$, D. Subramaniam$^3$,
  J. Burgd\"orfer$^2$ and M. Morgenstern$^3$}
\affiliation{ $^{1}$Dep. of Mechanical and Aerospace
  Engineering, Princeton University, Princeton, NJ 08544, USA\\$^{2}$Institute for Theoretical
  Physics, Vienna University of Technology, A-1040 Vienna, Austria, EU\\$^{3}$II. Institute of Physics B and JARA-FIT, RWTH Aachen, 52074 Aachen, Germany, EU}

\date{\today}

\begin{abstract}
The surface potential of the herringbone reconstruction on Au(111) is
known to guide surface-state electrons along the potential
channels. Surprisingly, we find by scanning tunneling spectroscopy
that hot electrons with kinetic energies twenty times larger than the
potential amplitude (38 meV) are still guided. The efficiency even
increases with kinetic energy, which is reproduced by a tight binding
calculation taking the known reconstruction potential and strain into
account. The guiding is explained by diffraction at the inhomogeneous
electrostatic potential and strain distribution
provided by the reconstruction.
\end{abstract}

\maketitle

\section{introduction}

Electromagnetic waves can be steered phase-co\-he\-rent\-ly along
interfaces of different dielectric con\-stant.\cite{Ghatak} An
analogous process of wave guiding is not well established for
ballistic electrons, e.g. in nanostructures.  Instead,
hard-wall potentials, \cite{Beenakker, vanWees, Song, McNeil} or
magnetic fields,\cite{Baranger, Geim97} which confine the electron
path classically, are invoked. However, exploiting the wave character
of the electrons for steering might be less invasive for the phase
in\-for\-ma\-tion with obvious consequences for the transport of
entangled quantum information.\cite{Beth,Hansel} Here, we probe the
model system Au(111), for which the well-known herringbone
reconstruction\cite{Barth90} provides a low-amplitude ($\VHB=38$ meV)
piecewise straight channeling potential with typical channel length of
$\lHB\approx 30$\,nm.\cite{Crommie98,Kern02}  The channeling
potential is periodic transverse to the channel direction with period
$\lambdaHB= 6.3$\,nm. Previous scanning tunneling microscopy (STM)
revealed standing electron waves preferentially along the channeling
potential at a kinetic energy of $\Ek:=E-\EG\approx 500$ meV ($\EG =
-480$\,meV, bottom of the surface band measured relative to the bulk
Fermi level $E_{\rm F}$), which implies guiding of hot surface
electrons.\cite{Fujita97}  However, a satisfactory explanation has
remained elusive: subsequent determination of the small amplitude of
the reconstruction potential,\cite{Crommie98,Kern02} with $\Ek/\VHB
\gtrsim 10$ rule out the originally proposed model\cite{Fujita97,
  Fujita973} of guiding by channeling in the potential well. Indeed,
the channeling interpretation has been challenged by showing that the
Talbot effect arising at an undulated step edge reveals a similar
standing wave pattern.\cite{Wenderoth} On the other hand, anisotropic
standing wave patterns have not been found for surface states on
unreconstructed close-packed surfaces as Cu(111), Ag(111), Pt(111) or
Ni(111).\cite{Petersen98,Jeandupeux99,Wiebe05,Braun08}

Here, we present a novel explanation of the guiding at high energies:
the regular dislocation lines act as a diffractive grating, leading to
a zeroth-order diffraction peak in the direction along the
reconstruction lines, and thus to diffractive focusing in momentum
space. Using STM and scanning tunneling spectroscopy (STS), we probe
the Au (111) surface state and identify two energy regimes: at low
energies, $\Ek \lesssim \VHB$, electrons channel within the hexagonal
close-packed (hcp) regions of the reconstruction as originally
proposed. At higher energies up to $\Ek/\VHB > 20$, the electrons
still propagate anisotropically, i.e. preferentially along the
reconstruction lines, but in all regions of the sample and not only in
the hcp regions. Diffraction explains both the large energy
$\Ek/\VHB\gg 1$ of steered electrons and the absence of a preferential
stacking area for wave guiding. 

 To uncover the ingredients of the
guiding effect, we perform tight binding (TB) simulations of the local
density of states (LDOS). Quantitative agreement between simulations
and experimental data requires to include the measured electrostatic
potential of the surface reconstruction,\cite{Kern02} the change in
effective mass due to the known compressive strain at the dislocation
lines,\cite{DFT_GOLD1} and disorder scattering at point defects.
Without strain, the guiding effect would still be present but weaker than in
the experiment.  Finally, we present a semiclassical model for the
interference of different electron paths subject to reflection at the
reconstruction lines and impurities. This model qualitatively reproduces the
effect, further corroborating that diffraction is, indeed, the origin
of the guiding.  Similar partial dislocation lines exist in other
structures, e.g., multilayer graphene.\cite{Ping, Hattendorf,Alden}
Our present findings for Au(111) may therefore open the door to
non-dispersive wave guiding for electrons by diffraction in such
structures.

\section{Experiment}

\begin{figure} [t]
\includegraphics[width= \linewidth ]{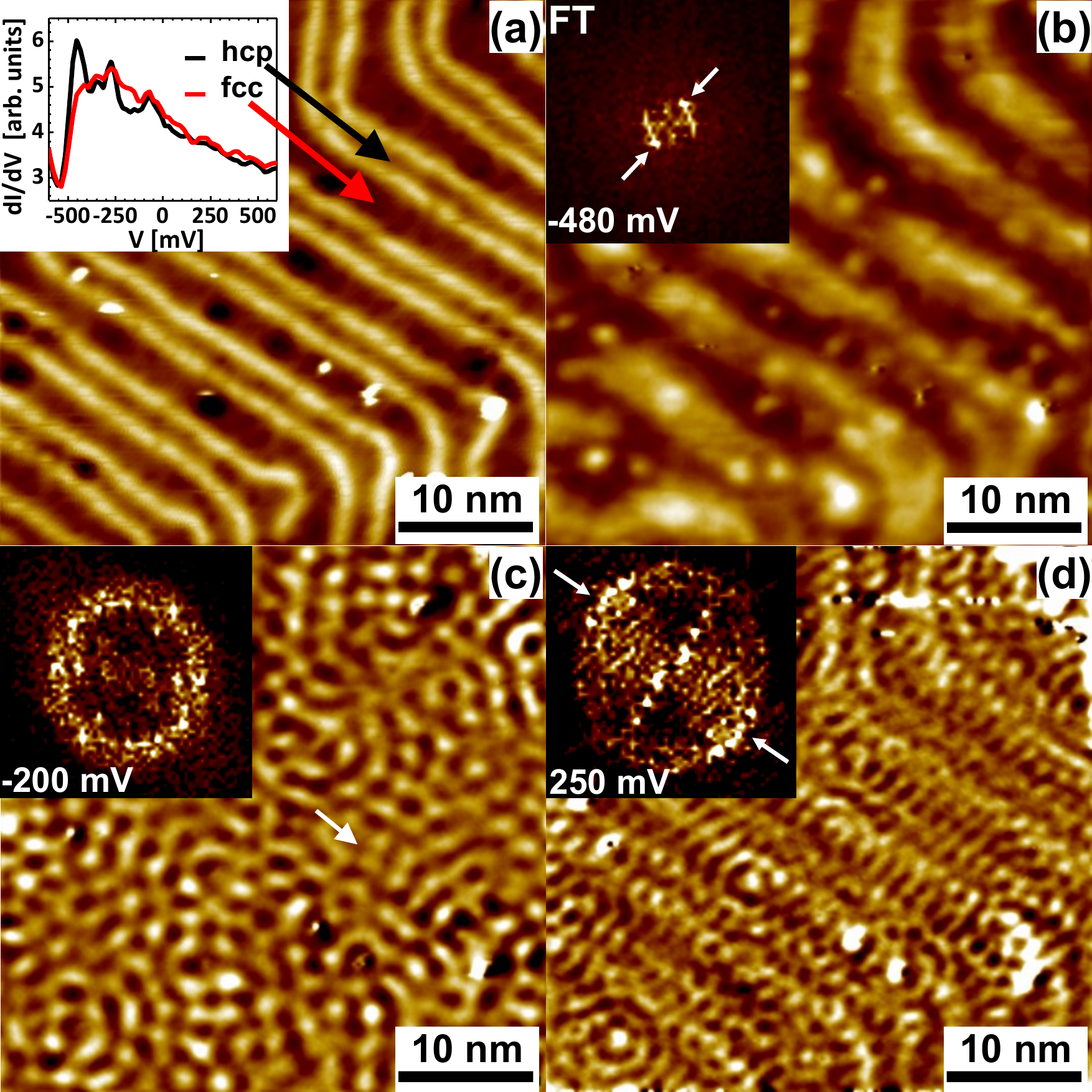}
\caption{(color online)
(a) STM image of Au(111), $I=1\rm\,nA$,
  $V=-480\rm\,mV$; herringbone reconstruction and a few impurities
  (bright and dark spots) are visible. Inset: Spatially averaged $dI/dV$
  spectra taken on hcp and fcc areas as indicated by arrows, $I_{\rm
    stab}=1\rm\,nA$, $V_{\rm stab}=-540$ mV, $V_{\rm
    mod}=10\rm\,mV$. (b)-(d) $dI/dV$ maps of the same surface area as
  in (a) recorded in constant-current mode at voltages as marked in the insets, $I=1\rm\,nA$, $V_{\rm
    mod}=10\rm\,mV$. Arrow in (c) indicates a guided standing wave. Insets: Fourier transformations (FT) of the $dI/dV$
  images; arrows in (b) mark spots corresponding to the herringbone periodicity; arrows in (d) mark the enhanced intensity within the circle
  in the direction parallel to the  reconstruction lines.
}
\label{fig1}
\end{figure}

The STM/STS experiments are performed
in ultrahigh vacuum at 5\,K.\cite{Mashoff09} The Au(111) surface is
prepared by cycles of Ar ion sputtering ($p_\mathrm{Ar}=1-3\cdot
10^{-5}\rm\,mbar$, $E_\mathrm{ion}=700\rm\,eV$) and annealing
($T=450\,^\circ$C, $\sim 10$\,min).\cite{Barth90, Crommie98,
  Wenderoth} After the preparation, the sample is transferred into
the precooled STM. STM images and $dI/dV$ maps, the latter recorded by
lock-in-technique with modulation voltage $V_{\rm mod}$, are taken in
constant-current mode at current $I$ and sample voltage $V$. The
$dI/dV$ curves are recorded with open feedback after stabilizing the
tip at current $I_{\rm stab}$ and voltage $V_{\rm stab}$.

\begin{figure}[t]
        \includegraphics[width= 1.0\linewidth]{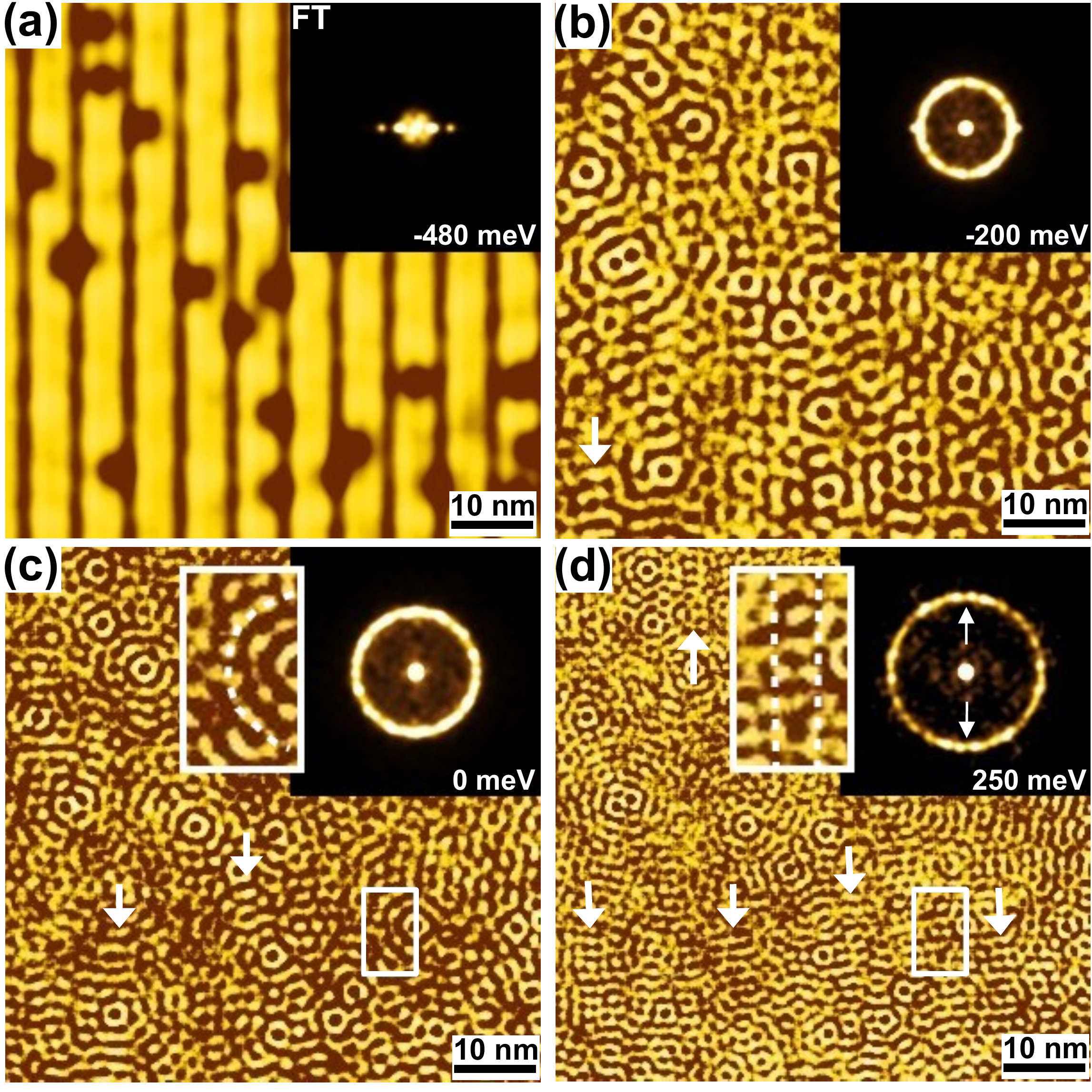}
\caption{(color online) Calculated LDOS maps at $E-E_{\rm F}$ as
  marked in the subfigures; reconstruction lines run vertically;
  insets in upper right corner are FTs of the real space images;
  arrows in LDOS images mark standing waves running along the
  reconstruction lines; pairs of white rectangles in (c) and (d) show
  a region close to a point scatterer and the corresponding zoom-in:
  wave guiding is visible at 250 mV, but not at 0 mV; arrows in FT
  point to the increased intensity along the reconstruction; the TB
  calculation includes the electrostatic potential and the strain of
  the reconstruction lines as well as impurities modeled as Gaussian
  potentials with amplitude 2.5 eV and width 0.3 nm.}
\label{fig2}
\end{figure}

STM images of Au(111) [Fig.~\ref{fig1}(a)] exhibit the well-known
$23\times \sqrt{3}$ herringbone reconstruction consisting of
alternating regions of hexagonal close-packed (hcp) and face-centered
cubic (fcc) stacking separated by bright partial dislocation
lines.\cite{Woell89, Barth90, Wenderoth} Additionally, a few
impurities are visible as bright spots. The $dI/dV$ curves
[Fig.~\ref{fig1} (a), inset] reproduce the well-known differences in
the electronic structure between hcp and fcc areas.\cite{Crommie98}
FTs of $dI/dV$ maps recorded at different $V$ in steps of $50\,$mV
[see Appendix~\ref{app:gallery} for a gallery] reveal the well-known
parabolic dispersion of the Au(111) surface state\cite{Kevan,LaShell}
with the origin at $\EG = -480$ meV and an effective mass $m^\star
\simeq 0.25 m_{\mathrm {e}}$ [as shown in
  Ref.~\onlinecite{GeringerPhD}], in good agreement with previous
photoemission\cite{Kevan} as well as scanning tunneling
spectroscopy\cite{Schouteden,Kern02} results. We also reproduce the
multiple phase shifts of the intensity distribution perpendicular to
the reconstruction lines which indicate miniband
formation.\cite{Didiot07, Didiot072, Didiot10,GeringerPhD} Here, we focus on the
anisotropy of standing waves visible in the $dI/dV$ maps [see
  Fig.~\ref{fig1}]. The higher $dI/dV$ intensity at $-480\,$mV in hcp
areas [Fig.~\ref{fig1}(b)] is caused by localization of surface
electrons in regions of lower electrostatic potential.\cite{Fujita97,
  Crommie98, Kern02} The 2D Fourier transformation (FT) of the $dI/dV$
map shows only spots related to the periodicity of the surface
reconstruction [Fig.~\ref{fig1}(b), inset]. The $dI/dV$ maps between
-400 mV and -150 mV corresponding to $\Ek > \VHB$ reveal ringlike
standing electron waves around point-like scatterers [see
  Fig.~\ref{fig1}(c)] as expected for isotropically delocalized
electrons.\cite{Heller94, Crommie93,Wittneven} In between,
interference patterns due to the random superposition of plane waves
appear as typical for 2D scattering.\cite{Morg02} The FT, accordingly,
shows a largely isotropic ring-like distribution of contributing
electronic wave vectors $\mathbf{k}$ [Fig.~\ref{fig1}(c), inset].

At higher energies ($V \gtrsim -150\,$mV, $\Ek \gtrsim 330\,$meV)
[Fig.~\ref{fig1}(d)], the electron standing waves become increasingly
oriented along the direction of the herringbone channels and much less
in the transverse direction.  Hence, the FT exhibits a ring with
anisotropic intensity being largest in the direction parallel to the
reconstruction lines as marked by arrows. The standing waves along the
reconstruction exist within the fcc and hcp regions which excludes
potential channeling as the origin of the guiding effect
[Appendix~\ref{app:corrogation}].

\section{Simulation}

To uncover the physical origin of guiding, we simulate the electronic
structure of the Au (111) surface band using a tight-binding
description in the continuum limit.\cite{MUMPS} We include the herringbone
reconstruction as a one-dimensional periodic on-site potential
\begin{eqnarray}\label{eq:Vpot}
W &=& \frac \VHB2 \left[1+\cos^2\left(\frac 12 \kHB x\right)\right]\cos^2(\kHB x) 
\label{eq:Vpot1}
\end{eqnarray}
with $\VHB = $38 $\mathrm{meV}$ the height of the reconstruction
potential and $k_{\rm HB}=2\pi/6.3$ nm$^{-1}$ its fundamental wave
vector as measured by Ref.~\onlinecite{Kern02}.  In order to include
the known compressive lateral strain variation within the
reconstruction,\cite{DFT_GOLD1} we perform ab-initio density
functional theory calculations of unreconstructed Au(111) with varying
lattice spacings. We use the VASP software
package,\cite{KresseFuerthau} including the associated PAW potential
for gold, and the PBE XC functional.\cite{PBE} We perform a geometry
relaxation of a 1x1 fcc (111) surface slab containing 11 layers
[i.e.~11 atoms in the unit cell] in a unit cell containing
20~\AA\ vacuum between the periodic images of the layer. For $k$-point
sampling in the slab, we use a Monkhorst-Pack grid of $18\times
18\times 1$. We relax the three top and bottom layers. Finally, we
calculate for the optimized geometry the surface band structure, and
identify the gold (111) surface state.  We deduce the
$m^\star$-dependence on strain by fitting the resulting parabolic
dispersion $\Ek = \hbar^2 {\bf k}^2/(2m^\star)$ ($\hbar$: Planck's
constant). From the literature \cite{DFT_GOLD1,DFT_GOLD2} the lattice
constant within the reconstructed top layer is known to vary around
the equilibrium lattice constant with the same period as the
reconstruction: the functional form is similar to the variation in the
on-site potential, Eq.~(\ref{eq:Vpot}), with a strain amplitude of
about 2\%. Accordingly, we investigate the variations in the
dispersion relation of the surface state as a function of the
stretching of the fcc lattice in the (100) direction up to 2\%. We
find a variation of the lower band edge of the order of 40 meV, in
agreement with the amplitude of the on-site potential of
Eq.~(\ref{eq:Vpot}). We also find a variation in the effective mass
$m^\star$ [in the (100) direction] of the surface state of $4\%$ for
the largest strain of $2\%$.  Such a change in effective mass
corresponds to a variation in the hopping parameter and
the on-site element of our tight-binding simulation to obtain the
correct continuum limit.

We account for disorder by including randomly distributed point-like
scattering potentials with a concentration $n\approx 0.01\,/{\rm
  nm}^2$ as taken from the STM measurements.  We use narrow Gaussian
peaks of height 2.5 eV and width $W\approx 3$ \AA. The results were
found to only weakly depend on the exact shape of the scattering
potentials, as long as their characteristic length scale is short
compared to the electron wavelength. Scattering by phonons and
electron-electron interaction has been found to exhibit a mean free
path $l_{\rm MFP}$ exceeding 30 nm in the investigated energy range
and is thus of minor importance.\cite{Reinert01, Echenique04,
  Campillo00} The finite length of the herringbone channels $\lHB$
does not explicitly enter the calculation except for the requirement
that the length $L$ of the patch calculated satisfies
$L\gtrsim\lHB\gg\lambdaHB$. We use an approximately quadratic patch of
150$\times$150 nm$^2$ with randomly shaped, soft-walled boundaries
(roughness amplitude $12\,$nm) to avoid unphysically preferred
directions due to boundary effects.  The LDOS of the patch is deduced
by summing the calculated electronic eigenstates over an energy window
of 25 meV (about 250 eigenstates at $\Ek=-0.2$ eV), in line with the
experimental energy resolution.  Superposing the squared eigenstates
yields an LDOS at $E$.  Fig.~\ref{fig2} shows LDOS maps and
corresponding FTs of the central area ($65 \times 65$ nm$^2$) of a
particular patch (for a gallery see Appendix A. Fig. 7).  The
calculated maps resemble the experimental LDOS maps in
Fig.~\ref{fig1}(b)-(d): channeling within the hcp regions at low
energy [Fig.~\ref{fig2}(a)], mostly isotropic scattering with some
regions of wave guiding (arrows) at intermediate energies
[Fig.~\ref{fig2}(b)(c)] and increasingly preferential standing waves
along the reconstruction lines at higher energy
[Fig.~\ref{fig2}(d)]. The FTs exhibit rings with angular anisotropy
being most obvious at the highest energy (arrows). The maximum FT
intensity is in the direction along the reconstruction lines.  A zoom
into an area close to a point scatterer [see white rectangles in
  Fig.~\ref{fig2}(c) and (d)] provides a real-space visualization for
the increase of guiding with energy: the standing waves encircle the
nearest defect in (c), while the ladder-like interference pattern in
(d) is formed by a standing wave running along the reconstruction
lines.

\begin{figure} [t]
        \includegraphics[width= \linewidth]{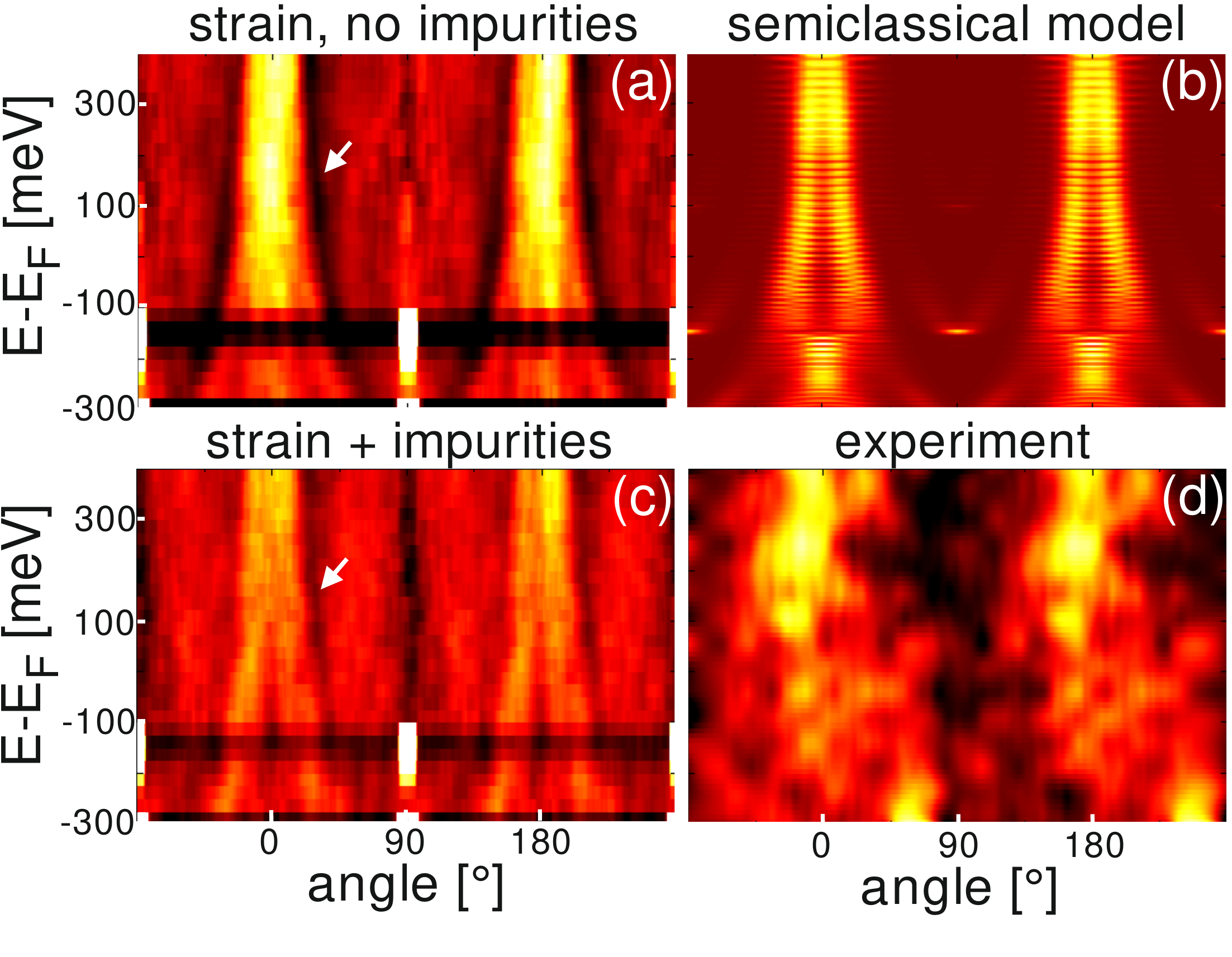}
\caption{(color online) Colorplot of the angular distribution of the
  FT averaged around the dominant $|{\mathbf k}|$ (ring structure in
  FTs of Fig.~\ref{fig1} and \ref{fig2}, respectively,
  Fig.~\ref{Figmaps} and \ref{fig_theor_crops}: (a) FT from ideal
  theoretical LDOS including reconstruction potential and strain but
  no disorder; (b) angular and energy dependence of backscattering
  probability according to a semiclassical Gutzwiller path calculation
  of the LDOS including reconstruction potential, strain and
  impurities (random phase shift of wave at each impurity $|\beta_R| <
  0.05\pi$); (c) same as (a) but with impurities as in Fig.~\ref{fig2}
  and Fig.~\ref{fig_theor_crops}; (d) FT from experimental LDOS for
  one domain of the herringbone reconstruction after angular smoothing
  (2nd order polynomial fit) and setting $eV=E-E_{\rm F}$. The color
  scale in all images covers the range [0.52, 1.00]; arrows at the
  dark lines in (a),(c) mark the dominant diffractive minimum
  [Eq.~(\ref{eq:diffmin})]; all theoretical FTs are averaged over 50
  realizations of disorder and different edge configurations for each
  energy, after subtracting the LDOS of a calculation with the same
  boundary, but without reconstruction or impurities, in order to
  remove the remaining influence of the boundary.  }
\label{fig3}
\end{figure}
 
\section{Comparison between experiment and simulation}

In order to quantify the correspondence between STS data and TB data,
we consider the angular-resolved FT intensity around the dominant
$|\mathbf{k}|$ (ring in the insets of Fig.~\ref{fig1} and \ref{fig2})
as a function of energy: FT$(\alpha,E)$ [Fig.~\ref{fig3}].  To remove
any residual numerical artifacts from the grid or boundary directions
within the calculation, we subtract for each patch the results from an
identical calculation without any reconstruction potential or point
scatterers. The subsequent average over 50 realizations further
minimizes remaining interference effects between waves scattered from
the boundary and from the reconstruction potential and impurities
within the FT$(\alpha, E)$. TB calculations without [Fig
  \ref{fig3}(a)] and with impurities [Fig \ref{fig3}(c)] show a clear
focusing effect, visible as a bright stripe around $0^\circ$ and
$180^\circ$. The bright stripe gets sharper and more intense with
increasing energy.  Such an increase is expected for diffraction,
since smaller wavelength (higher energy) decreases the angle of
constructive interference for a given spacing of the
grating. Consequently, a focusing around the forward direction in
$k$-space appears. The maxima are, indeed, delimited by a
stripe (arrows) indicating the lowest order diffractive
minimum of the dominant Fourier component $2\kHB$ of the
reconstruction potential at
\begin{equation}\label{eq:diffmin}
\sin \alpha=\frac{\hbar\kHB}{2\sqrt{2\Ek m^\star}}.
\end{equation}
Without impurities the interference structure is more pronounced,
as expected. 

\begin{figure}[h]
        \includegraphics[width= \linewidth]{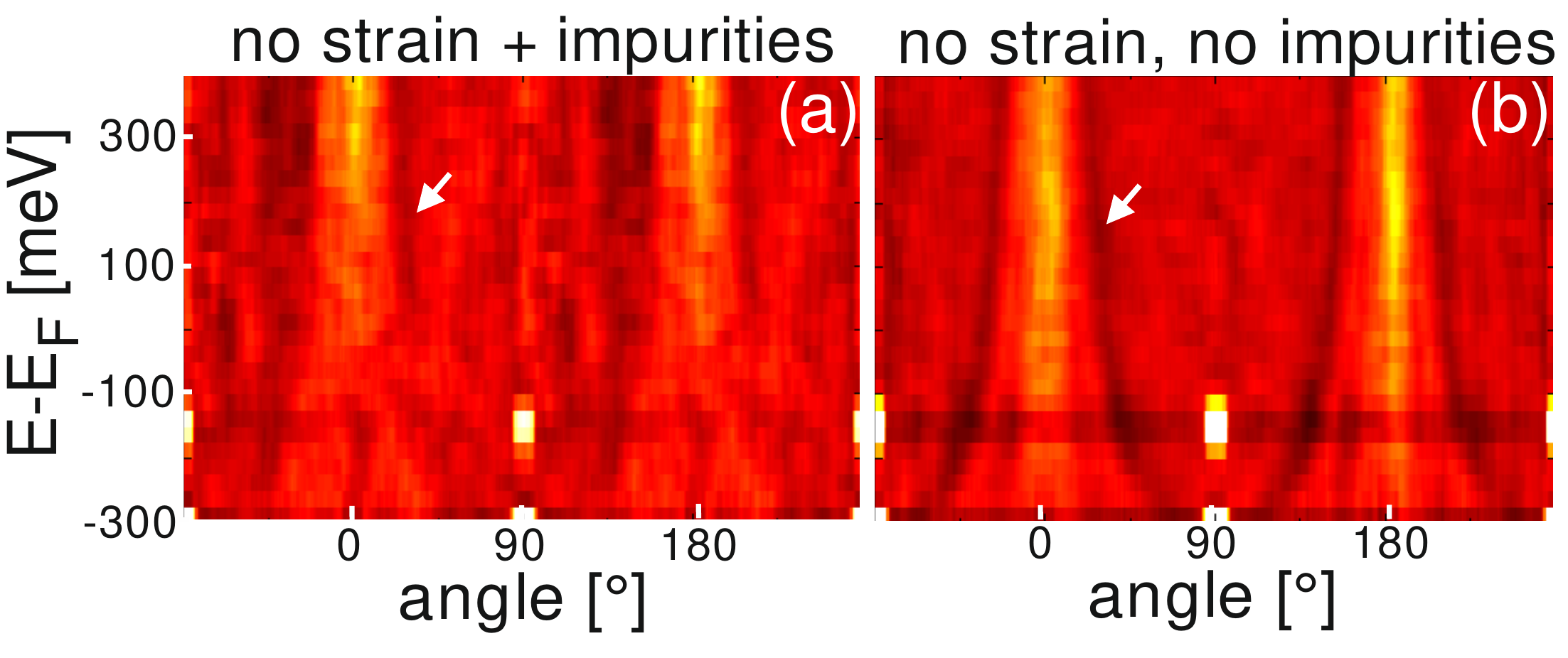}
\caption{(color online) Angular dependence of the FT of theoretical
  LDOS maps, but in contrast to Fig.~\ref{fig3} without considering
  any strain effects (same color scale and contrast as in
  Fig.~\ref{fig3}). (a) with and (b) without disorder scattering.}
\label{fig3NS}
\end{figure}

To elucidate the conditions required for the observation of electron
wave guiding along the direction of the herringbone reconstruction
pattern, we also perform a semiclassical simulation. We
consider the interference pattern created by superposition of closed
semiclassical orbits. We use the complex reflection and transmission
amplitudes at a single period of the reconstruction potential to
evaluate the weight and phase of semiclassical closed orbits based on
a Gutzwiller trace formula.\cite{Brack} The semiclassical action
along each path $C$ is given by
\begin{equation}
S = \int_C k dq + \mu + \beta_R.
\end{equation}
The phase $\mu\left(t(k),r(k)\right)$ accounts for the phase shifts
accumulated by reflections at the reconstruction potential (similar to
a Maslov index), and the (randomly determined) small angle $\beta_R
\in [-\beta_{\max}(k),+\beta_{\max}(k)]$ accounts for random phases
accumulated by impurity scattering. The weight of each path is
determined by the accumulated reflection/transmission probabilities
along the path. We include all combinations of multiple reflections at
the periodic potential, for up to 16 internal reflections, until
numerical convergence is reached. We use Gutzwiller's trace formula to
derive, from the superposed periodic orbits, the local density of
states.\cite{Brack} Upon summation of the different complex amplitudes
associated with paths for a given starting angle $\alpha_0$ we obtain
the angular distribution as a function of energy. We obtain the
increased guiding of electrons at higher energies along the direction
of the reconstruction lines [Fig.~\ref{fig3}(b)]. Moreover, our model
qualitatively reproduces (i) the dependence of the onset of wave
guiding on the amount of dephasing by impurity scattering
$\beta_{\max}$ and (ii) the dependence of the onset of wave guiding on
the maximal allowed path length (limited in experiment by both the
inelastic mean free path, and the length of the herringbone
channel). Indeed, for high disorder concentration, the only remaining
discernible feature is the zeroth-order diffraction peak. Due to the
drastic approximations made in terms of modeling of impurity
scattering as random phases and multiple scattering events at the
herringbone potential, we do not expect exact quantitative agreement.
Nevertheless, our semiclassical model supports the notion that the observed
channeling is, indeed, a direct consequence of diffractive scattering
of waves on the scattering grid created by the herringbone potential.
We conclude that interference, i.e. diffraction, accounts for the
effect.

The experimental FT$(\alpha,E)$ [Fig.~\ref{fig3}(d)] shows striking
similarities to TB calculations and the semiclassical model, i.e. a
bright stripes around $0^\circ$ and $180^\circ$, which get stronger
and sharper with increasing $\Ek$. The reduced sharpness in the
experiment is most likely due to angular smoothing required for noise
suppression.  Since the appearance of a pronounced diffractive minimum
required averaging over 50 disorder realizations in the simulation for
each energy, we do not expect to see it for a single series of
experimental $dI/dV$ images.  Note that the contrast of the
experimental pattern is quantitatively similar to the calculated ones
[Fig.~\ref{fig3}]. Finally, calculations without considering the
change in effective mass due to strain [see Fig.~\ref{fig3NS}] show
qualitatively similar features as with strain, yet feature a weaker
contrast than the experiments and the calculations including
strain. We thus conclude that strain is decisive to reproduce the
strength of the guiding effect.

\begin{figure} [t]
        \includegraphics[width= 1\linewidth]{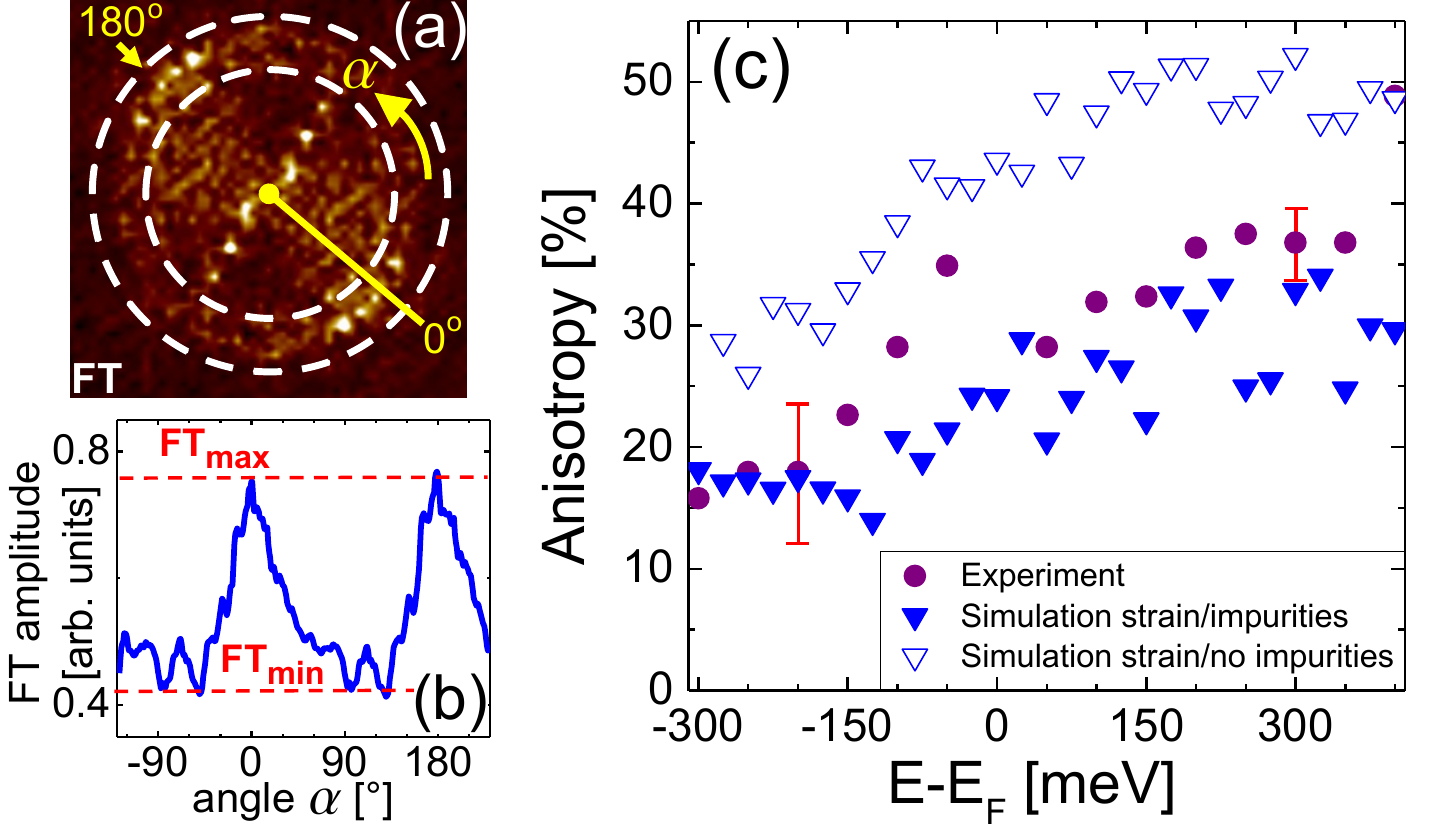}
        \caption{(color online) (a) FT of $dI/dV$ map at $V=250\rm\,V$
          recorded within a single domain of the reconstruction lines;
          dashed rings border the area used for the angular plot in
          (b); angles of maximum intensity $\alpha=0\,^\circ$ and
          $\alpha=180\,^\circ$ are marked. (b) Angular distribution of
          FT amplitude radially averaged over the ring marked in (a);
          minimum FT$_{\rm min}$ and maximum FT$_{\rm max}$ are
          marked. (c) Anisotropy $A$ according to Eq.~(\ref{eq:anis})
          from experimental data [circles] and from TB simulations
          ($\triangledown$: without, $\blacktriangledown$: with scatterers). Experimental
          error bars decrease with energy as indicated.}
\label{fig4}
\end{figure}

Figure \ref{fig4} shows a direct quantitative comparison between
experiment and TB simulation.  We cut out the ring area of the FT
belonging to standing waves [Fig.~\ref{fig4}(a)], which results in the
angular distribution FT$(\alpha)$ [Fig.~\ref{fig4}(b)].
FT$(\alpha)$ displays pronounced maxima at the angles along the
reconstruction lines. We define the anisotropy $A$ by
\begin{equation}
A:=\frac{\mathrm{FT}_{\max} - \mathrm{FT}_{\min}}{\mathrm{FT}_{\max}+
  \mathrm{FT}_{\min}}. \label{eq:anis}
\end{equation}
with $\mathrm{FT}_{\max}$ and $\mathrm{FT}_{\min}$ being the maximum
and the minimum of $\mathrm{FT}(\alpha)$,
respectively. Figure~\ref{fig4}(c) reveals that the experimental data
for $A$ increase with energy from $A<0.2$ at $E-E_{\rm F}\simeq-300$
mV to $A> 0.35$ for $E-E_{\rm F}\simeq 350$ mV.  The TB simulation
without impurities reproduces the increase, but with absolute values
of $A$ being larger than in the experiment. Including impurities
reduces $A$ to reach a reasonable agreement with the experimental
data.  Neglecting the strain yields a strongly reduced anisotropy that
only weakly increases with energy. Even without impurities we obtain
$A=0.15$ at $E-E_{\rm F}\simeq -300$~mV and $A=0.2$ at $E-E_{\rm
  F}\simeq 400~$mV, i.e. values much lower than in experiment and
barely increasing with energy.  This emphasizes that strain
contributes significantly to the guiding as can be rationalized by the
influence of effective mass ($m^\star$) and on-site potential ($W$) on
the parabolic energy dispersion. In other words, strain changes the
curvature of a parabolic band, while the electrostatic potential only
changes the origin of the parabola on the energy axis.  Consequently,
the distance of k-points at high energy is much larger in case of
strain than in case of electrostatic potential. This implies a higher
reflection coefficient in case of strain and, thus, a stronger guiding
by the reconstruction, explaining the increase of intesity in
FT$(\alpha,E)$ in forward direction with increasing energy (Fig.~3).\\

\section{Conclusion}

In conclusion, we have presented a joint experimental and theoretical
analysis of guiding of surface electrons on Au (111) by the
herringbone reconstruction. We identify two energy regimes.  At low
energies, electrons assemble in the hcp channels of the reconstruction
potential implying channeling.  At higher energies above the
confinement potential, the angular focusing into the direction of the
channels is induced by diffraction. This focusing even increases with
energy due to a locally varying effective mass induced by the strain
distribution in the reconstruction. Remarkably, the guiding parallel
to the herringbone channels persists up to energies that exceed the
height of the reconstruction potential by a factor of 20.

\acknowledgments

We gratefully acknowledge helpful discussions with M. Liebmann and U. Klemradt and
support by the Max Kade Foundation NY, and
the SFB-041 VICOM. Numerical calculations were performed on the Vienna
Scientific Clusters 1+2.

\appendix
\section{Comparison of measured and calculated LDOS images on Au(111)}
\label{app:gallery}

\begin{figure*}[t]
\includegraphics[width=0.85\textwidth]{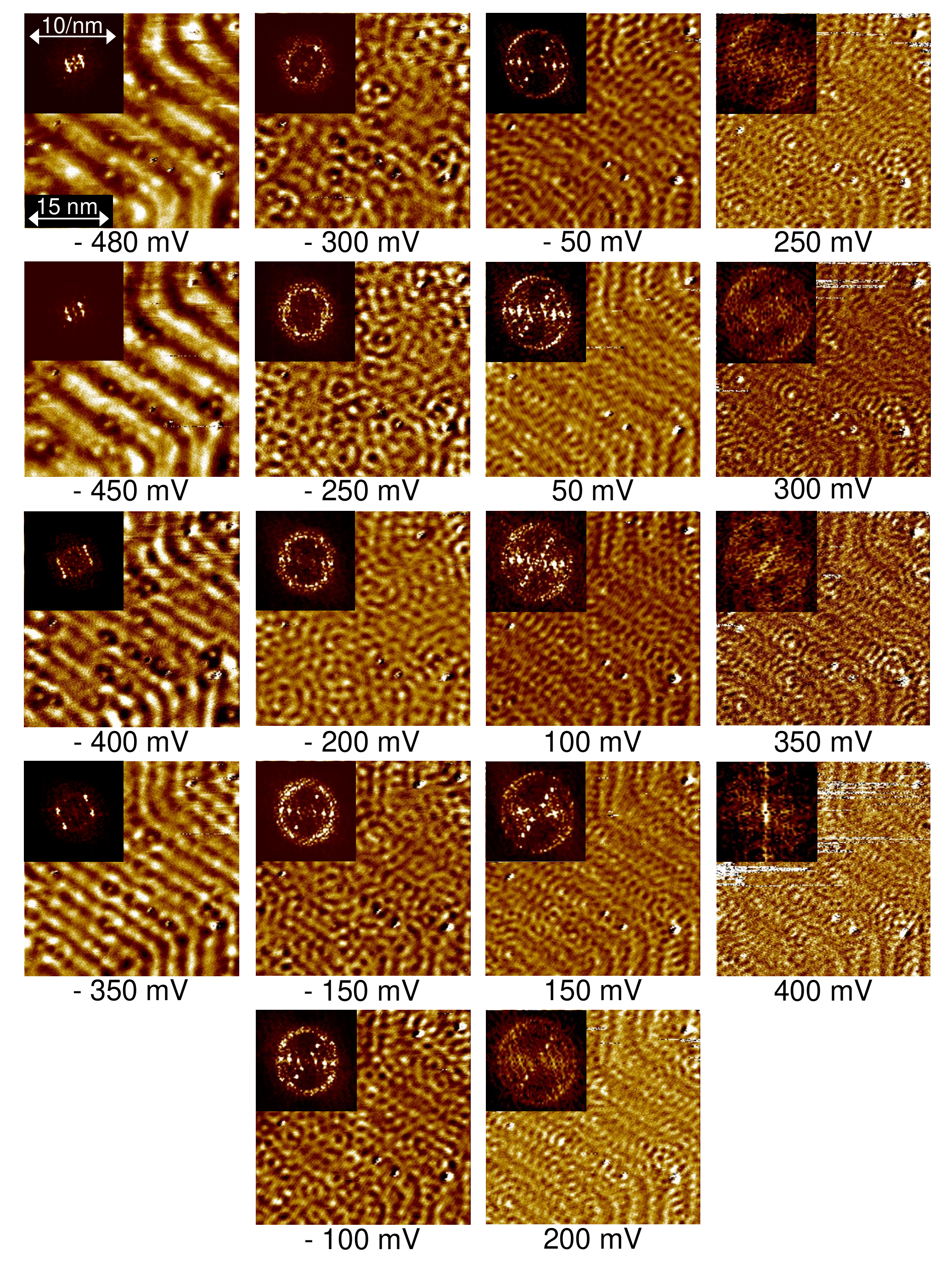}
\caption{\label{Figmaps} $dI/dV$ maps of the same surface area at
  voltages $V$ as marked below the images, $I=1\rm\,nA$, $V_{\rm
    mod}=10\rm\,mV$; Insets: Fourier transformations of the $dI/dV$
  images; size of $dI/dV$ map and Fourier transformation are the same
  for all $V$; contrast is adapted to improve visibility.}
\end{figure*}

\begin{figure*}[t]
\includegraphics[width=0.85\textwidth]{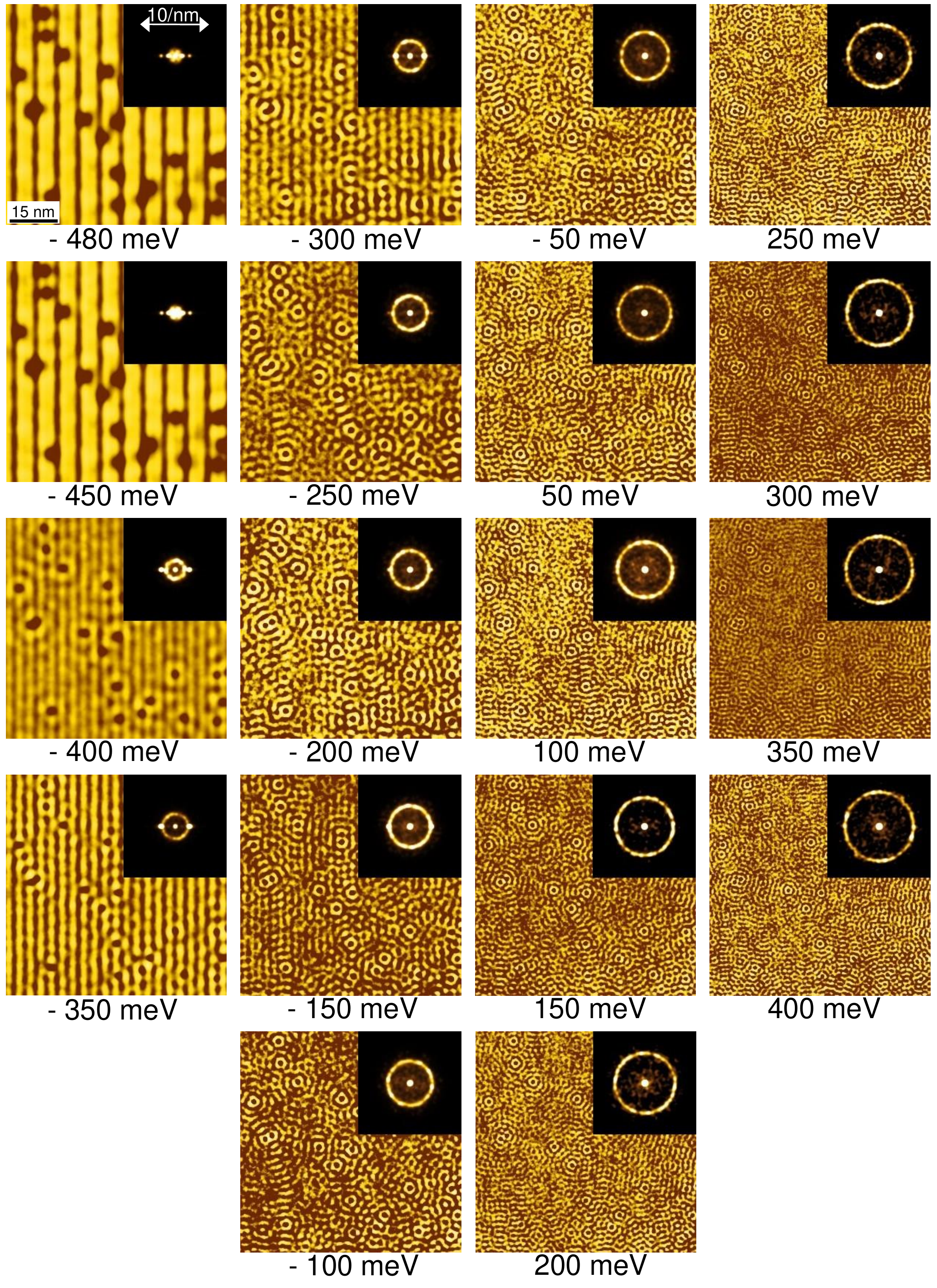}
\caption{\label{fig_theor_crops} Theoretical LDOS maps at voltages $V$
  as marked below the images for a particular configuration of
  impurities, compare with the experimental LDOS in
  Fig.~\ref{Figmaps}. The reconstruction lines run from top to
  bottom. Insets: Fourier transformations of the $dI/dV$ images; size
  of $dI/dV$ map and Fourier transformation are the same for all $V$;
  contrast is adapted to improve visibility.}
\end{figure*}

Figs.~\ref{Figmaps} shows experimental LDOS images of the same area of
a Au(111) surface recorded at various energies (see
insets). Fig.~\ref{fig_theor_crops} shows the calculated LDOS patterns
for comparison. The insets show the corresponding Fourier
transformations (FTs). The images feature slightly varying contrast in
order to improve visibility. Features in the experimental and
theoretical data are largely identical, albeit the strength of the
features coming directly from the reconstruction lines appear more
intense in the experiment. At low voltage the confined states with
nodes and antinodes perpendicular to the reconstruction lines dominate
the LDOS map. Consequently, two pairs (one pair) of spots are visible
in the experimental (calculated) FT. In experiment, the two pairs
correspond to the two different domains visible in the main
image. Between $-450$ and $-400$ mV, a doubling of antinodes is visible
in the main image leading to a doubling of the distance between the
pairs of spots in the experimental FT.  
The theoretical LDOS also
features faint spots at larger distance and energies -480 and -450
meV. \clearpage Notice that confinement of the electrons within the
reconstruction potential of height 38 meV \cite{Crommie98, Kern02} is only
present within the images recorded at $-480$ mV and $-450$ mV.
At $-400$ mV and $-350$ mV, i.e. at energies
above the confinement potential, first circular structures are visible
around defects, which are attributed to standing electron waves.
 They
lead to a faint circle within the FT exhibiting an increasing radius
with increasing energy.
Between $-300$ mV and $-200$ mV, this circle becomes more intense with
respect to the spots from the reconstruction showing that the
influence of the confinement potential gets weaker. The intensity
within the circle is rather homogeneous indicating largely isotropic
scattering of electron waves. This is also visible within the LDOS
images, where no direction appears to be preferred by the electron
waves.  Above $-150$ mV, the two lines of spots within the circle,
representing nodes and antinodes perpendicular to the reconstruction
lines, get more intense again. In addition, the circular intensity
within the FT becomes asymmetric with larger intensity in the upper
left and the lower right region in experiment, and in the upper and
lower region within the simulations. This brighter region is caused by
the steering of the electron waves along the reconstruction lines,
which is visible in the real space LDOS images too.  The brighter area within
the ring is composed of intensity perpendicular to the lines of spots
originating from the reconstruction. To demonstrate this, the
experimental FTs above $200$ mV are not taken from the whole image,
but only from the area where reconstruction lines run diagonally
through the image. Then, only one line of spots and the corresponding
perpendicular bright areas within the ring appear. Thus, the
transition from confined states at low energy to largely isotropic
scattering at medium energies, and to a continuously increasing
wave guiding at higher energies is clearly discernible in theory and
experiment.

\begin{figure}[t]
\includegraphics[width=0.95\linewidth]{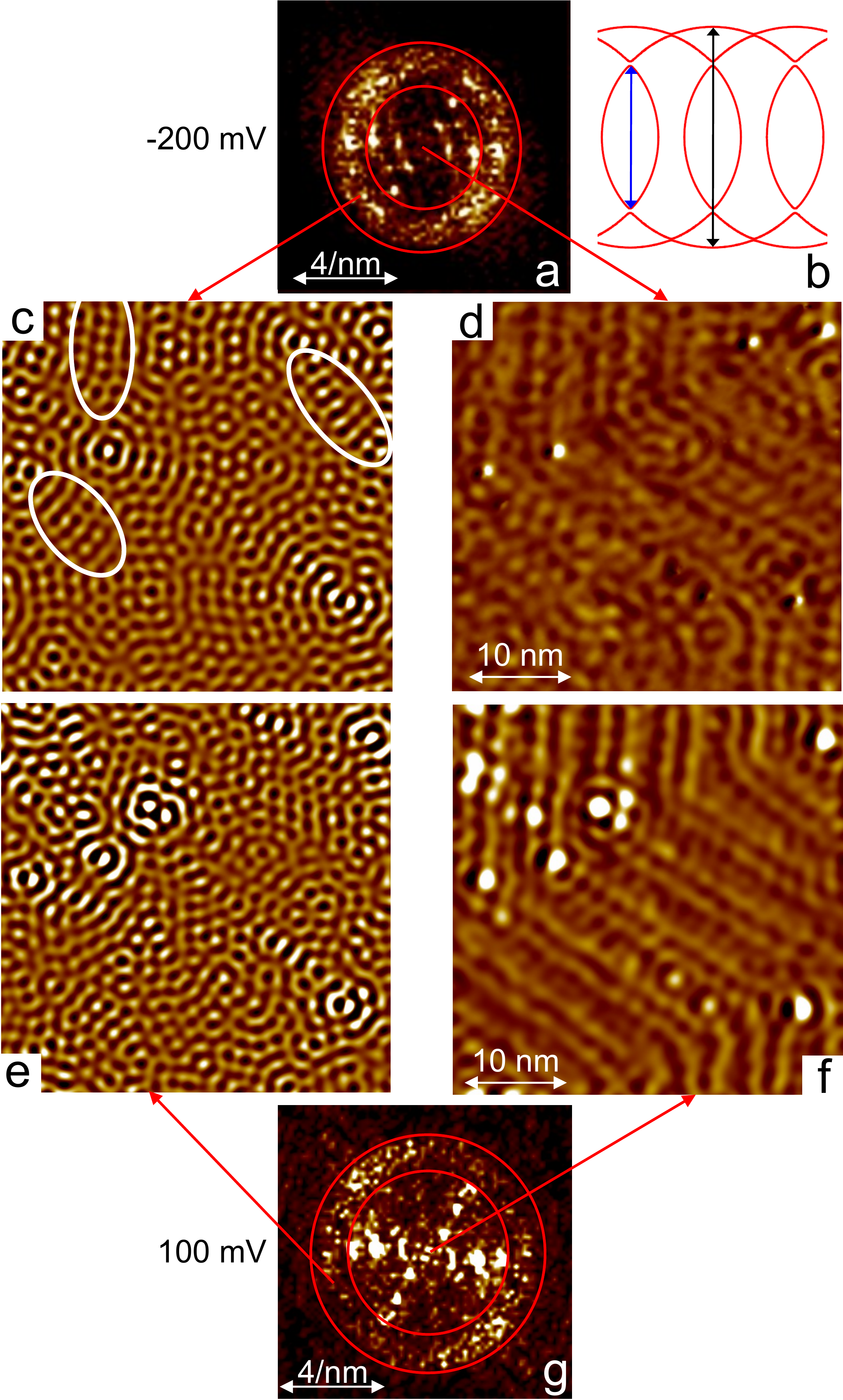}
\caption{\label{Fig5} (a) Power spectrum of Fourier transformation of
  $dI/dV$ map recorded at $V=-200$ mV; regions for back-transformation
  in (c) and (d) are marked by red circles; (b) five backfolded
  constant-energy circles in $\mathbf{k}$ space for Au(111) with gaps
  originating from miniband formation and marked preferential
  scattering vectors parallel to the reconstruction lines (arrows);
  (c) Image resulting from inverse Fourier transformation of (a) after
  applying an ideal band-pass filter selecting the ring structure,
  which is bordered by the two red circles in (a); white circles mark
  areas of wave guiding; (d) Image resulting from inverse Fourier
  transformation of (a) after applying an ideal low-pass filter
  selecting the area bordered by the inner red circle; (e), (f) same
  as (c),(d) but for a $dI/dV$ image recorded at $V=100$ mV and
  backtransformed from (g); (g) Power spectrum of Fourier
  transformation of $dI/dV$ map recorded at $V=100$ mV; regions for
  back-transformation in (e) and (f) are marked by red circles.}
\end{figure}

\begin{figure}[t]
        \includegraphics[width= \linewidth]{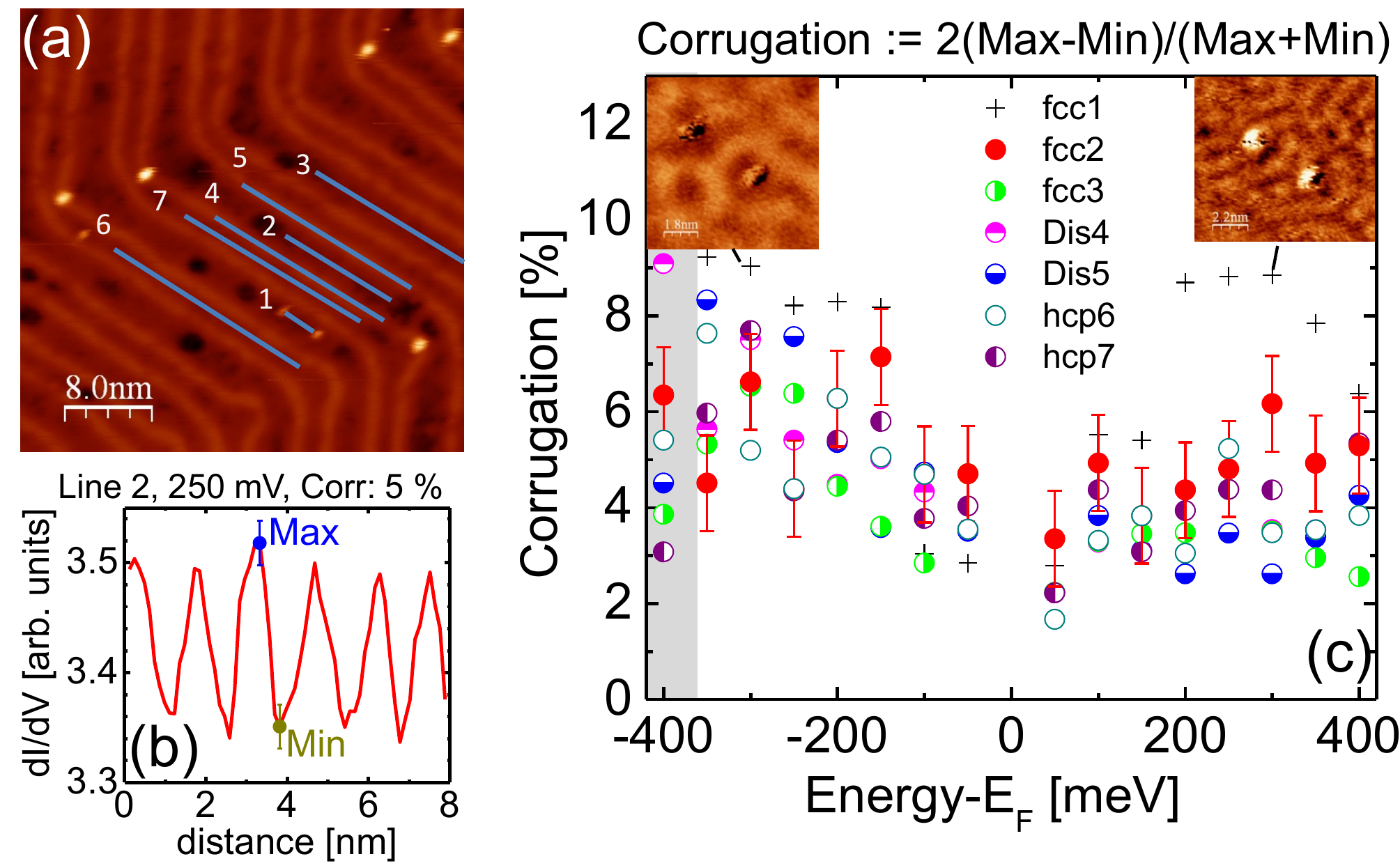}
\caption{(a) STM image of Au(111), $I=1$ nA, $V=-0.48$ V with marked
  lines where the corrugation $C$ is measured. (b) $dI/dV$ intensity
  along line 2 in (a) recorded at $V=0.25$ V; minima (Min) and maxima
  (Max) as used for the evaluation of $C$ are marked. (c) Corrugation
  $C$ as determined by Eq.~(\ref{eq:corrugation}) as a function of
  electron energy with respect to $E_{\rm F}$ for the lines marked in
  (a); labels include the type of region (Dis = dislocation line) as
  well as the number marking the lines in (a); Insets show $dI/dV$
  images corresponding to the marked points of relatively large
  corrugation; grey area marks the channeling energies; $I=1$ nA,
  $V_{\rm mod}=10$ mV for all data points and the insets.}
\label{Fig4}
\end{figure}
A more detailed analysis of the contributions to the FT
[Fig.~\ref{Fig5}] proves that the ring within the power spectrum of
the FT is, indeed, caused by standing electron waves while the
intensity spots at lower $\mathbf{k}$ are mostly due to the Au(111)
reconstruction.  To disentangle contributions of the guided electrons
and the herringbone reconstruction, we partition the FT into the ring
area around the Fermi circle, and its interior [see two red circles in
  in Fig.~\ref{Fig5}(a)]. We consider energies $V=-200$ mV
[Fig.~\ref{Fig5} (a)] and $V=100$ mV [Fig.~\ref{Fig5}(g)]. Both the
ring and the circle are back-transformed separately into real-space
images as indicated by arrows. The backtransformed images of the ring
area are shown in Fig.~\ref{Fig5}(c) and (e) and the backtransformed
images of the inner circle area are shown in Fig.~\ref{Fig5}(d) and
(f).  The $dI/dV$ images corresponding to the ring nicely show
electron wave patterns which are regularly distributed over the
surface.  This regular distribution is typical for 2D states because
of the larger probability for closed trajectories of the electrons on
a path including several defects.\cite{Morg02,Morg}\\ It is obvious that most
of the electron wave patterns are isotropic in Fig.~\ref{Fig5}(b), but
that wave guiding along the reconstruction lines already exists in
certain areas of the sample as marked by white rings. This explains
that the anisotropy shown in Fig.~\ref{fig3}(c) does not drop
to zero at lower energy.  In Fig.~\ref{Fig5}(e) the wave guiding is
evidently stronger. Notice that wave guided patterns of electron waves
are also observed in Fig.~\ref{Fig5}(f) and (d). We believe, that this
is due to the fact that band gaps open at the crossing points of
different constant-energy circles (the original one and backfolded
ones) as shown in Fig.~\ref{Fig5}(b), which implies additional
scattering vectors along the reconstruction lines but with shorter
wave vectors (blue arrow) than for scattering across the original
circle (black arrow).

\section{Corrugation}
\label{app:corrogation} 

We determine the corrugation of the standing electron waves along the
reconstruction line for different perpendicular coordinates labeled by
1 to 7 in Fig.~\ref{Fig4}(a).  Lines 1 to 3 are located in fcc
regions, lines 4 and 5 on top of dislocation lines, and 6 and 7 in
hcp regions. For each section line in $dI/dV$ images as shown, e.g. in
Fig.~\ref{Fig4}(b), the maxima (Max) and minima (Min) determine the relative
corrugation $C$ according to
\begin{equation}
\label{eq:corrugation}
  C(E) = 2\frac{(dI/dV)_{\max} - (dI/dV)_{\min}}{(dI/dV)_{\max}+(dI/dV)_{\min}}.
\end{equation}
$C(E)$ is found to be largely independent of the stacking and, thus,
of the local reconstruction potential [Fig.~\ref{Fig4}(c)], i.e. the
observed steering is spatially homogeneous. This further corroborates
that the steering effect is not a real-space channeling in regions of
minimal potential. $C(E)$ depends also only weakly on energy, except
where Fabry-Perot resonances between adjacent scatterers appear [see
  e.g.~Line 1 in Fig.~\ref{Fig4} connecting two adjacent defects]. Two
resonances with larger $C(E)$ appear if one or two antinodes between
the two defects are observed [see insets].

\end{document}